\documentclass[aps,preprint,showpacs,preprintnumbers,amsmath,amssymb]{revtex4}
\usepackage{amsmath,mathrsfs,amsbsy,color,graphicx,bm,amsthm,amsfonts}
\usepackage{units}
\usepackage{bbm}
\usepackage{times}
\usepackage{dcolumn}
\usepackage{mathrsfs}
\usepackage{amsmath,amssymb,epsfig}
\usepackage{amsmath}
\newcommand{\udots}{\mathinner{\mskip1mu\raise1pt\vbox{\kern7pt\hbox{.}}
\mskip2mu\raise4pt\hbox{.}\mskip2mu\raise7pt\hbox{.}\mskip1mu}}
\begin{document}
\title{Genuine multipartite entanglement subject to the Unruh and anti-Unruh
effects}
\author{Shu-Min Wu$^1$\footnote{smwu@lnnu.edu.cn}, Hao-Sheng Zeng$^2$\footnote{Corresponding author: hszeng@hunnu.edu.cn}, Tonghua Liu$^3$\footnote{liutongh@mail.bnu.edu.cn}}
\affiliation{$^1$ Department of Physics, Liaoning Normal University, Dalian 116029, China\\
$^2$ Department of Physics, Hunan Normal University, Changsha 410081, China\\
$^3$ Department of Physics, Yangtze University, Jingzhou 434023, China}


\begin{abstract}
We study the acceleration effect on the genuine tripartite entanglement for one or two accelerated detector(s) coupled to the vacuum field. Surprisingly, we find that the increase and decrease in entanglement have no definite correspondence with the Unruh and anti-Unruh effects. Specifically, Unruh effect can not only decrease but also enhance the tripartite entanglement between detectors; Also, anti-Unruh effect can not only enhance but also decrease the tripartite entanglement. We give an explanation of this phenomenon. Finally, we extend the discussion from tripartite to N-partite systems.
\end{abstract}

\vspace*{0.5cm}
\maketitle
\section{Introduction}
Hawking in 1974 discovered black hole evaporation caused by thermal radiation emitted by the black hole \cite{L3}. Soon after, Unruh in 1976 proposed that the Minkowski vacuum observed by an inertial observer would, as seen by an accelerated observer, be detected as a thermal bath of particles, which means that the content of particles is observer dependent \cite{L1,L2}. The Unruh effect has not been confirmed experimentally so far, but some detection schemes have been proposed \cite{LK1,LK2,LK3}.  On the other hand, the Hawking radiation has also made great progress in experimental simulation \cite{LK4,LK5,LK6}.
The two phenomena are closely related and the study of Unruh effect is helpful to understand the Hawking radiation, and to study the thermodynamics and the problem of information loss \cite{L4,L5,L6}.
The Unruh effect and Hawking radiation have been extended to different fields, such as free scalar field,  free Dirac field and so on \cite{L7,L8,L9,LL9,LMA9,LMA10,L10,L11,L12,L13,L14,L15,L16,L17,L19,LK8,LK9}. In general, quantum steering, quantum entanglement and quantum discord between quantum fields decrease with the increase of observer acceleration, which means that Unruh effect is harmful to quantum resources based on quantum fields.

Recently, another interesting phenomenon, the so-called anti-Unruh effect, has been discovered. It says that a Unruh-DeWitt detector (a two-level quantum system interacting with the quanta of the external field at a particular frequency) accelerating uniformly in the Minkowski vacuum may become less excited at higher accelerations than at lower \cite{L20,L23}.
This phenomenon is contrary to the common sense that a two-level atom embedded in a thermal bath would be excited, and is named as the so-called anti-Unruh effect.
Though the physical essence about anti-Unruh effect has not been understood completely, some relevant issues, such as the influence of anti-Unruh effect on the various types of quantum resource, deserve to be investigated. These investigations are important for relativistic quantum information science and its applications. Currently, the action of anti-Unruh effect on the quantum coherence, entanglement and phase sensitivity have been examined \cite{L21,L22}. In Reference \cite{L21}, the authors considered two causally separated qubits (Unruh-DeWitt detectors) in an entangled initial state, where each qubit independently accelerates in its respective vacuum cavity. They showed that anti-Unruh effect can enhance the entanglement of the qubit systems. Contrary to previous view, we show both the Unruh effect and anti-Unruh effect can enhance or decrease the entanglement between arbitrary numbers of detectors. 
It has been shown previously that acceleration can either enhance or degrade the harvested entanglement that can be harvested in a two-detector system from the vacuum state in an Unruh-DeWitt detector's model \cite{LAZ22,LAZ23,LAZ24}. However, here we will consider an initially entangled state in our model.

As the information task becomes more and more complex, more entangled particles are needed to deal with relativistic quantum information. Therefore, we consider an entangled N-partite Unruh-DeWitt detectors. But for the sake of clarity, we start with the entangled tripartite systems. Assume three observers, Alice, Bob and Charlie, each of them holds a point-like Unruh-Dewitt detector, which interacts locally with its scalar fields. The three detectors are initially in a tripartite entangled state in the Minkowski spacetime. When one or two of the observers accelerate uniformly, the tripartite entanglement between detectors will change. Our motivation is to show how the Unruh effect and anti-Unruh effect influence the tripartite entanglement, and whether new interesting phenomena can be found. After this, we extend the tripartite system to the N-partite systems and discuss similar problems.

The paper is organized as follows. In Sec. II, we briefly recall the quantification of the multipartite entanglement. In Sec. III, we briefly introduce the Unruh-DeWitt detector model.  In Sec. IV, we study the changes of the tripartite entanglement under both the Unruh and anti-Unruh effects. In Sec. V, we extend the issues from tripartite to N-partite entangled systems.
The last section is devoted to the conclusion and discussion.

\section{Quantification of genuine multipartite entanglement \label{GSCDGE}}
Multipartite entanglement is defined by its opposite of biseparability.
We call a N-partite pure state $|\Psi\rangle\in H_1\otimes H_2\otimes...\otimes H_N$ to be biseparable,
if it can be written in the form $|\Psi\rangle=|\Psi_A\rangle\otimes|\Psi_B\rangle$, with
$|\Psi_A\rangle\in H_A=H_{i_1}\otimes H_{i_2}\otimes...\otimes H_{i_k}$ and
$|\Psi_B\rangle\in H_B=H_{i_{k+1}}\otimes H_{i_{k+2}}\otimes...\otimes H_{i_N}$ in any  bipartition of the Hilbert space. Obviously, a biseparable pure state has at least one pure marginal. If $|\Psi\rangle$ is not biseparable with respect to its any bipartition, it is called genuine N-partite entanglement (For brevity, we omit the word ``genuine" in what follows and call it ``N-partite entanglement" in the following description).
The measure for the multipartite entanglement is defined as \cite{L26}
\begin{eqnarray}\label{Q1}
E(|\Psi\rangle)=\min_{\chi_i\in\chi}\sqrt{2[1-\text{Tr}(\rho^2_{A_{i}})]},
\end{eqnarray}
where $\chi=\{A_i|B_i\}$ denotes the set of all possible bipartitions of the whole $N$-partite system, and $\rho_{A_{i}}$ is the marginal state for the subsystem $A_{i}$.
The multipartite entanglement for the mixed state $\rho$ can be got by the convex roof construction
\begin{eqnarray}\label{Q2}
E(\rho)=\inf_{\{p_i,|\Psi_i\rangle\}}\sum_ip_iE(|\Psi_i\rangle),
\end{eqnarray}
where $\rho=\sum_ip_i|\Psi_i\rangle\langle\Psi_i|$ takes over all possible decompositions.

For N-qubit systems, the natural basis $\{|0,0,...,0\rangle,|0,0,...,1\rangle,...,|1,1,...,1\rangle\}$ are favorable. A N-qubit system is said to be in the X state if its density matrix in the natural basis can be written in form
\begin{eqnarray}\label{Q3}
 \rho_X= \left(\!\!\begin{array}{cccccccc}
a_1 &  &  &  &  &  &  & z_1\\
 & a_2 &  &  &  &  & z_2 & \\
 &  & \ddots &  &  &  \udots &  & \\
 &  &  & a_n & z_n &  &  & \\
 &  &  & z_n^* &  b_n & &  & \\
 &  & \udots &  &  & \ddots &  & 0\\
 & z_2^* &  &  &  &  & b_2 & \\
z_1^* &  &  &  &  &  &  & b_1
\end{array}\!\!\right),
\end{eqnarray}
where $n=2^{N-1}$.  The conditions $\sum_i(a_i+b_i)=1$ and $|z_i|\leq\sqrt{a_ib_i}$
are required, so that $\rho_X$ is normalized and positive. The N-qubit entanglement of a X state is given by \cite{L27}
\begin{eqnarray}\label{Q4}
E(\rho_X)=2\max \{0,|z_i|-\nu_i \},   i=1,\ldots,n,
\end{eqnarray}
where $\nu_i=\sum_{j\neq i}^n\sqrt{a_jb_j}$.

\section{The Unruh-DeWitt model \label{GSCDGE}}
In order to simulate the interaction between the quantum field and the detector, the Unruh-DeWitt model is usually used, which consists of a two-level quantum system or atom locally coupled to a scalar field obeying the Klein-Gordon equation along its trajectory \cite{L28}. Theoretically, the scalar field could be massive or massless. In this paper, we following the idea of reference \cite{L20} and assume that the scalar field is massless. Consider a cylindrical cavity with length $L$, which is initially in the Minkowski vacuum. A Unruh-DeWitt detector accelerates uniformly in the cavity along the length direction. The detector has its ground state $|g\rangle$ and excited state $|e\rangle$ which are separated by an energy gap $\Omega$. Due to
the Fulling-Unruh radiation, the detector actually feels a non-vacuum action of scalar field $\phi(x)$.
In the ($1 + 1$)-dimensional model, the action can be described in the interaction picture by the following Hamiltonian
\begin{eqnarray}\label{Q5}
H_I=\lambda\chi(\tau/\sigma)\mu(\tau)\phi(x(\tau)),
\end{eqnarray}
where $\lambda$ is the coupling strength that is assume to be weak, $\tau$ is the detector's proper time, $\mu(\tau)=e^{i\Omega\tau}\sigma^++e^{-i\Omega\tau}\sigma^-$ is the detector's monopole moment, and $\chi(\tau/\sigma)$ is the switching function which controls the duration of interaction via the parameter $\sigma$. As in most references, the switching function is taken as
\begin{eqnarray}\label{Q6}
\chi(\tau/\sigma)=e^{-\tau^2/2\sigma^2}.
\end{eqnarray}
For weak coupling ($\lambda\ll 1$), the corresponding unitary evolution for the whole quantum system can be written as
\begin{eqnarray}\label{Q7}
U=\mathbb{I}+U^{(1)}+\mathcal{O}(\lambda^2)=\mathbb{I}-i\int d\tau H_I(\tau)+\mathcal{O}(\lambda^2).
\end{eqnarray}
In the first-order approximation, the evolution could be further described as \cite{L20,L21,L22,L23}
\begin{eqnarray}\label{Q9}
U|g\rangle|0\rangle=\frac{1}{\sqrt{1+|\eta_0|^2}}(|g\rangle|0\rangle-i\eta_0|e\rangle|1\rangle_{g}), \nonumber
\end{eqnarray}
\begin{eqnarray}\label{Q9}
U|e\rangle|0\rangle=\frac{1}{\sqrt{1+|\eta_1|^2}}(|e\rangle|0\rangle+i\eta_1|g\rangle|1\rangle_{e}).
\end{eqnarray}
Where $|1\rangle_{g}$ denotes the normalized one-particle state of the scalar field produced by the evolution of state $|g\rangle|0\rangle$ under the unitary transformation $U$, and similar notation for the state $|1\rangle_{e}$. Formally, we can express $|1\rangle_{g}=\sum_{k}\xi_{k}|1_{k}\rangle$ and $|1\rangle_{e}=\sum_{k}\zeta_{k}|1_{k}\rangle$ with $\Sigma_{k}|\xi_{k}|^{2}=1$ and $\Sigma_{k}|\zeta_{k}|^{2}=1$.  The mode $k$ of the scalar field has annihilation and creation operators $a_k$ and  $a^\dag_k$ satisfying $a_k|0\rangle=0$ and $a_k^\dag|0\rangle=|1_k\rangle$. Obviously, for the multiple scalar field, the two states $|1\rangle_{g}$ and $|1\rangle_{e}$ are different in general. But for single mode scalar field, they are equal. In this paper, we continue to adopt the idea of references \cite{L21,L22} and make the single mode approximation in calculating the entanglement between detectors, i.e., set $|1\rangle_{g}=|1\rangle_{e}$. The periodic boundary condition leads to discrete modes $k=2\pi m/L$ with $m$ being integer.  $\eta_0=\lambda\sum_m I_{+,m}$ and $\eta_1=\lambda\sum_m I_{-,m}$ are associated with the excitation and deexcitation
probability of the particle, with $I_{\pm,m}$ given by
\begin{eqnarray}\label{Q10}
I_{\pm,m}=\int_{-\infty}^{\infty}\chi(\tau/\sigma) e^{\pm i\Omega\tau+\frac{2\pi i}{L}[|m|t(\tau)-mx(\tau)]} \frac{d\tau}{\sqrt{4\pi|m|}}.
\end{eqnarray}
In this paper, we assume that the accelerated trajectory of the detector is
$t(\tau)=a^{-1}\sinh(a\tau)$ and
$x(\tau)=a^{-1}[\cosh(a\tau)-1]$ with $a$ being the proper acceleration.

If the detector is initially in its ground state, then the excitation probability, at the leading order in the perturbative expansion, is given by
\begin{eqnarray}\label{QQ}
P=\sum_{m\neq0}|\langle1,e|U^{(1)}|0,g\rangle|^2=\lambda^2\sum_{m\neq0}|I_{+,m}|^2.
\end{eqnarray}
Generally, for the larger acceleration of the detector, the excitation probability of the detector is also larger. This is the so-called Unruh effect. Contrarily, if the excitation probability of the detector becomes less excited at higher accelerations than at lower, the phenomenon is named the anti-Unruh effect.

Garay et. al. systematically analyzed the condition for the emergence of anti-Unruh effect \cite{L23}. They showed that the existence of anti-Unruh effect depends on some form of low energy cut off in the external field. This could be due to either the field mass or the presence of an externally implemented infra-red (IR) cut off. There would be no anti-Unruh behaviour for massless fields without an externally implemented IR cut off. In the recent research about the property of Fulling-Unruh radiation, it was shown that the mass of the scalar field can lead to the non-Planckian corrections to the Fulling-Unruh radiation \cite{L35,L36,L37}. Indeed, it is this deviation of the radiation from  thermality that leads to the emergence of anti-Unruh effect \cite{L23}. This means that anti-Unruh effect is a kind of perception only for the accelerated detectors. Inertial detectors coupled to generic thermal baths would not perceive anti-Unruh effect. In this paper, we consider the massless scalar field, but adopt ``IR cut off", i.e., remove the constant mode ($m=0$) in the summation of Eq.(\ref{QQ}), because it would lead to divergency of Eq.(9). As we will see below, this implementation could lead to the emergence of anti-Unruh effect.

\section{The evolution of genuine tripartite entanglement \label{GSCDGE}}
In reference \cite{L21}, the authors considered an entangled state of two Unruh-DeWitt detectors, where either one detector is uniformly accelerated or both of them are accelerated simultaneously in their respective vacuum cavities. Then they studied how the anti-Unruh effect influences the quantum entanglement between detectors. Now we want to take one step further based on this model, by generalizing the systems from bipartite entanglement to tripartite entanglement.

We consider three causally separated cavity-detector systems. For convenience, we denote the detectors as A, B and C, held by Alice, Bob and Charlie respectively. The three cavity-detector systems are completely the same, except for their different space positions. Assume that the detectors are initially in an entangled Greenberger-Horne-Zeilinger-like (GHZ-like) state, and the cavities in a product vacuum state, i.e., the initial compound state is
\begin{eqnarray}\label{Q11}
|\psi\rangle_{ABC}=[\alpha|g_Ag_Bg_C\rangle+\beta|e_Ae_Be_C\rangle]|0_{A}0_{B}0_{C}\rangle,
\end{eqnarray}
where the nonzero $\alpha$ and $\beta$ are assumed to be real and satisfy $\alpha^2+\beta^2=1$.
In what follows, we will distinguish two cases according to the acceleration of detectors: (i) Alice and Bob remain stationary, while Charlie moves with a constant acceleration in his vacuum cavity; (ii) Alice remains stationary, while Bob and Charlie move with the same constant acceleration.

\subsection{Charlie accelerating}
Firstly, we consider the case (i), i.e., Alice and Bob stay stationarily,  while Charlie moves with a uniform acceleration in his vacuum cavity.
Applying the transformation of Eq.(\ref{Q9}) for detector C, we get
\begin{eqnarray}\label{Q12}
|\psi\rangle_{ABC_I}&=&\frac{\alpha}{\sqrt{1+|\eta_0|^2}}|g_Ag_Bg_C0_A0_B0_C\rangle-
\frac{i\alpha\eta_0}{\sqrt{1+|\eta_0|^2}}|g_Ag_Be_C0_A0_B1_C\rangle \\
&+&\frac{\beta}{\sqrt{1+|\eta_1|^2}}|e_Ae_Be_C0_A0_B0_C\rangle+
\frac{i\beta\eta_1}{\sqrt{1+|\eta_1|^2}}|e_Ae_Bg_C0_A0_B1_C\rangle. \nonumber
\end{eqnarray}
This implies that the vacuum cavity defined by inertial observers is not any more vacuum from the view of uniformly accelerated observers, and the detector C will feel the production of particles.

In order to study the tripartite entanglement between detectors, we trace the degrees of freedom over the field modes and obtain the reduced density matrix of the detectors
\begin{eqnarray}\label{Q13}
 \rho_{ABC_I}= \left(\!\!\begin{array}{cccccccc}
a_1 & 0 & 0 & 0 & 0 & 0 & 0 & z_1\\
 0 & a_2 & 0 & 0 & 0 & 0 & z_2& 0 \\
 0 & 0 & 0 & 0 & 0 &  0 & 0 & 0 \\
0 & 0 & 0 & 0 & 0 & 0 & 0 & 0 \\
0 & 0 & 0 & 0 &  0 & 0 & 0 & 0\\
0 & 0 & 0 & 0 & 0 & 0 & 0 & 0\\
0 &z_2^* & 0 & 0 & 0 & 0 & b_2 & 0\\
z_1^* & 0 & 0 & 0 & 0 & 0 & 0 & b_1
\end{array}\!\!\right),
\end{eqnarray}
where the density matrix is written in order of bases $|g_Ag_Bg_C\rangle$, $|g_Ag_Be_C\rangle$, $|g_Ae_Bg_C\rangle$, $|g_Ae_Be_C\rangle$, $|e_Ag_Bg_C\rangle$, $|e_Ag_Be_C\rangle$, $|e_Ae_Bg_C\rangle$ and $|e_Ae_Be_C\rangle$, and the nonzero elements are given by
\begin{eqnarray}\label{Q14}
&&a_1=\frac{\alpha^2}{1+|\eta_0|^2}, a_2=\frac{\alpha^2|\eta_0|^2}{1+|\eta_0|^2}, \\ \nonumber
&&b_1=\frac{\beta^2}{1+|\eta_1|^2}, b_2=\frac{\beta^2|\eta_1|^2}{1+|\eta_1|^2}, \\ \nonumber
&&z_1=\frac{\alpha\beta}{\sqrt{(1+|\eta_0|^2)(1+|\eta_1|^2)}},z_2=\frac{-\alpha\beta\eta_0\eta_1^*}{\sqrt{(1+|\eta_0|^2)(1+|\eta_1|^2)}}.
\end{eqnarray}
Substituting these parameters into Eq.(\ref{Q4}), we obtain the tripartite entanglement between detectors
\begin{eqnarray}\label{Q15}
E(\rho_{ABC_I})=\max \bigg\{0, \frac{2\alpha\beta(1-|\eta_0||\eta_1|)}{\sqrt{(1+|\eta_0|^2)(1+|\eta_1|^2)}},
\frac{2\alpha\beta(|\eta_0||\eta_1|-1)}{\sqrt{(1+|\eta_0|^2)(1+|\eta_1|^2)}}\bigg\}.
\end{eqnarray}
This expression suggests that the tripartite entanglement between detectors now depends not only on the initial state (parameters $\alpha$ and $\beta$), but also on acceleration $a$, the setup's parameters (energy gap $\Omega$ and cavity length $L$), and the interaction time $\sigma$.

\begin{figure}
\begin{minipage}[t]{0.5\linewidth}
\centering
\includegraphics[width=3.0in,height=5.2cm]{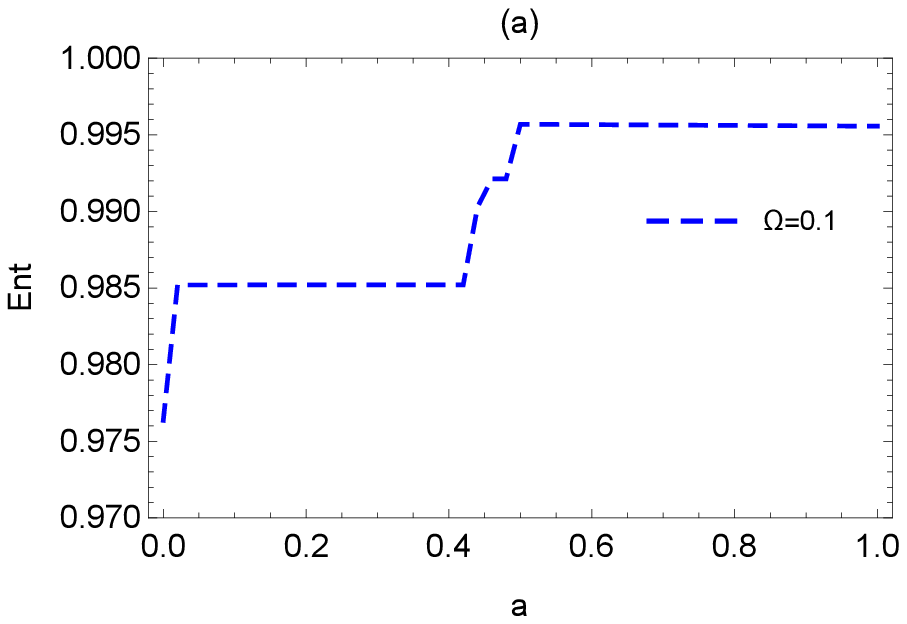}
\label{fig1a}
\end{minipage}%
\begin{minipage}[t]{0.5\linewidth}
\centering
\includegraphics[width=3.0in,height=5.2cm]{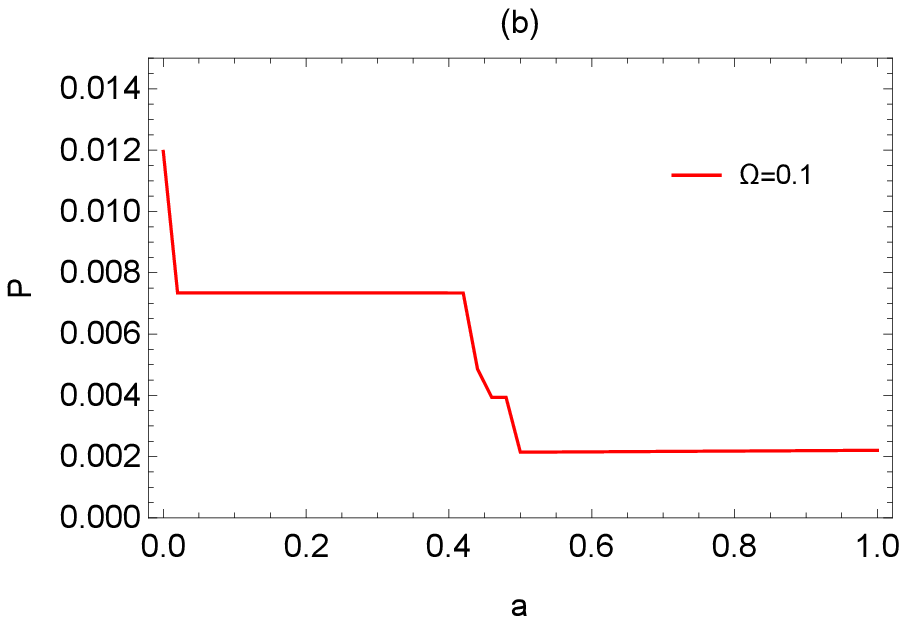}
\label{fig1b}
\end{minipage}%

\begin{minipage}[t]{0.5\linewidth}
\centering
\includegraphics[width=3.0in,height=5.2cm]{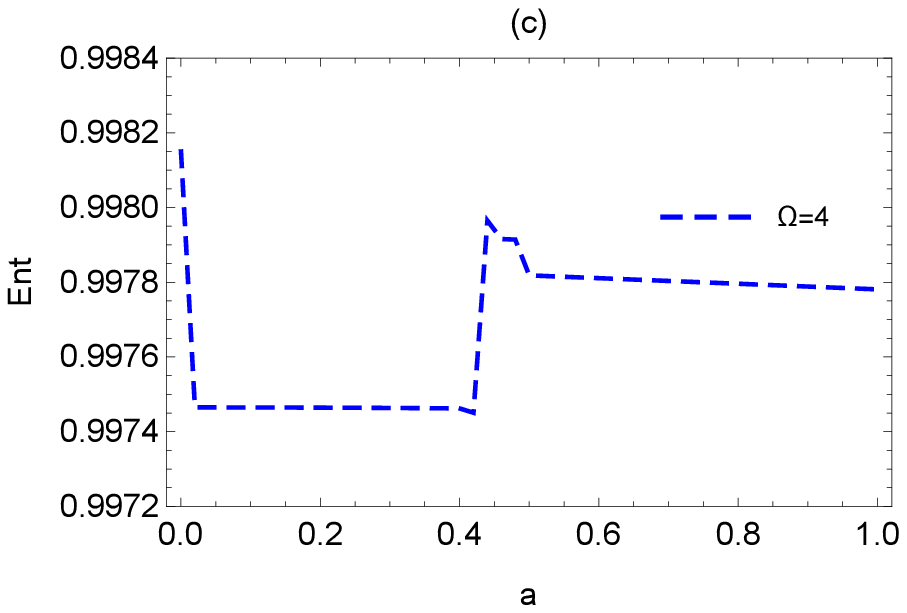}
\label{fig1c}
\end{minipage}%
\begin{minipage}[t]{0.5\linewidth}
\centering
\includegraphics[width=3.0in,height=5.2cm]{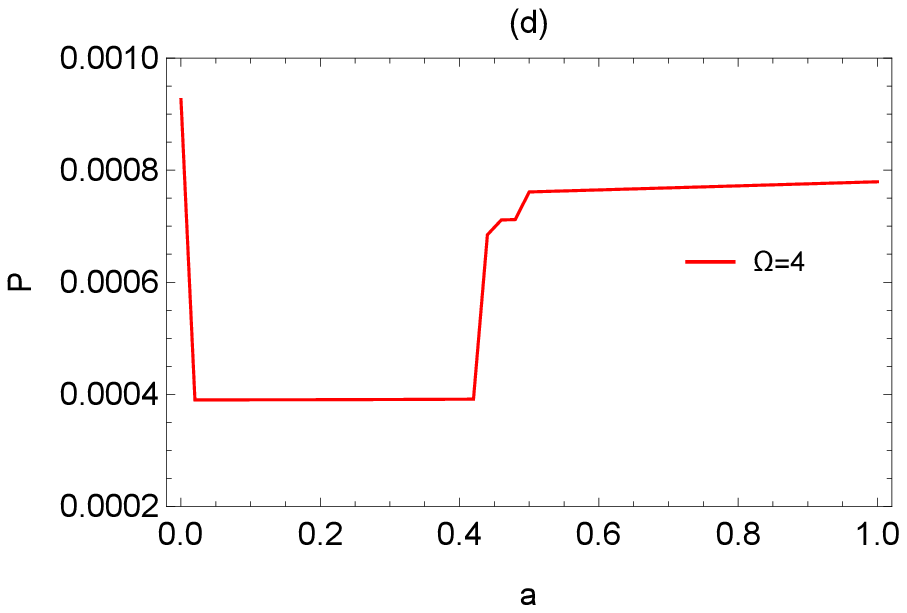}
\label{fig1d}
\end{minipage}%

\begin{minipage}[t]{0.5\linewidth}
\centering
\includegraphics[width=3.0in,height=5.2cm]{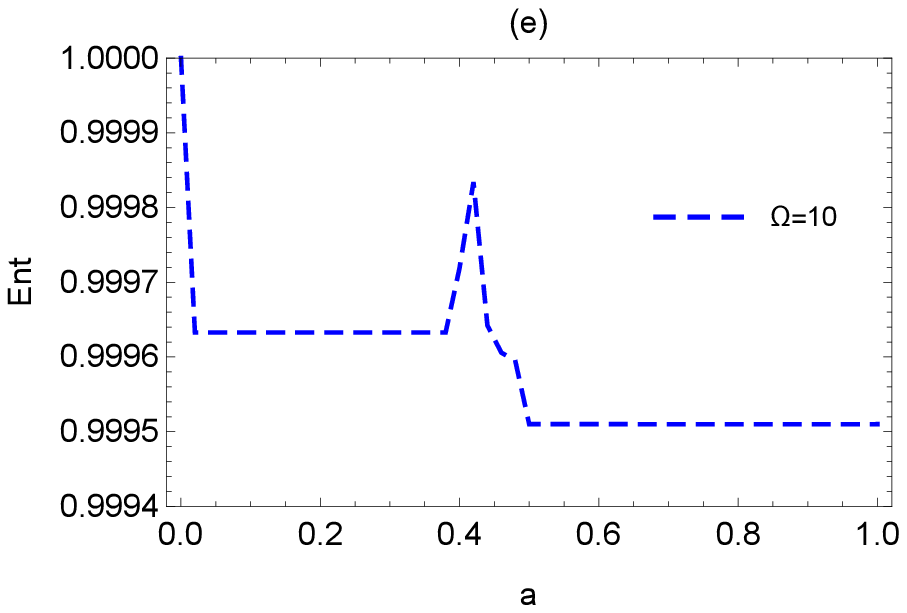}
\label{fig1e}
\end{minipage}%
\begin{minipage}[t]{0.5\linewidth}
\centering
\includegraphics[width=3.0in,height=5.2cm]{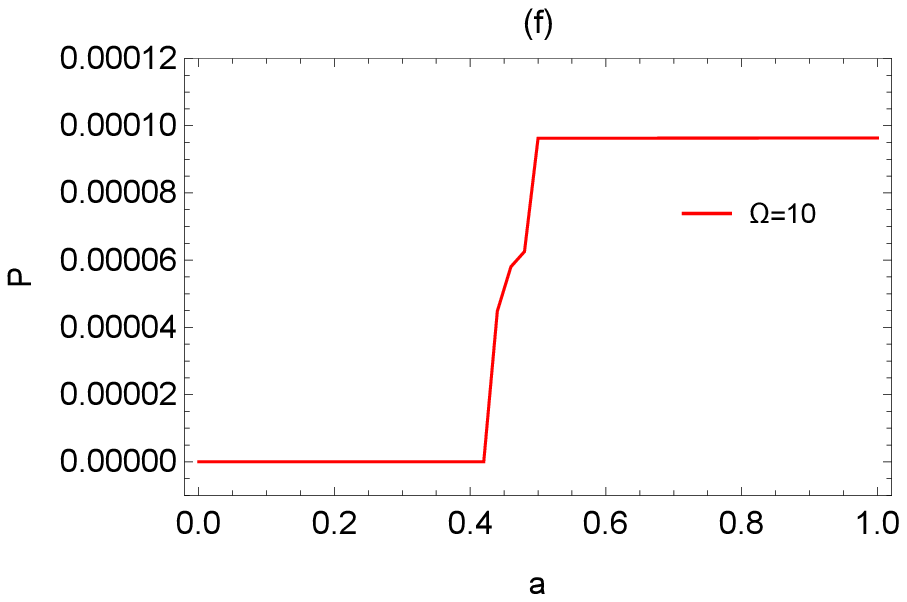}
\label{fig1f}
\end{minipage}%

\caption{Tripartite entanglement $E(\rho_{ABC_I})$ and the corresponding transition probability $P$ as functions of the acceleration $a$ for different the energy gap $\Omega$. The other parameters are fixed as $\sigma=0.4$, $L=200$ and $\alpha=\frac{1}{\sqrt{2}}$. }
\label{Fig1}
\end{figure}

In Fig.\ref{Fig1},  we plot the tripartite entanglement $E(\rho_{ABC_I})$ and the detector's transition probability $P$ given by Eq.(\ref{QQ}), for the smaller interaction time ($\sigma=0.4$), as functions of the acceleration $a$ for different energy gap $\Omega$.
Hereafter, the coupling strength is fixed as $\lambda=0.1$.
The top row in Fig.\ref{Fig1} shows that, with the increasing of the acceleration $a$, the entanglement increases and the corresponding transition probability $P$ decreases, meaning that anti-Unruh effect enhances the tripartite entanglement. This result is consistent with the case of bipartite entanglement in \cite{L21}. From the middle row in Fig.\ref{Fig1}, we see that the tripartite entanglement decreases for very small acceleration $a$ in which the corresponding transition probability $P$ also decreases, meaning that anti-Unruh effect can also decrease entanglement. For the parameter region about $a\in(0.42,0.45)$, the entanglement increases and the corresponding transition probability also increases, meaning that Unruh effect can enhance entanglement. These two results are counterintuitive which are just opposite to the previous result \cite{L21}. Finally from the last row in Fig.\ref{Fig1}, we see that, in the parameter region about $a\in(0.4,0.5)$, the transition probability increases, but the entanglement increases firstly and then decreases(appears a peak), meaning that Unruh effect can both enhance and decrease entanglement. Note that the entanglement for $a=0$ is generally less than one, which is due to the effect of the switching function. For the infinite interaction time, this phenomenon will disappear \cite{L21}. Physically, we can understand this phenomenon as a kind of entanglement transfer: When the detector C enters the cavity, it interacts with the cavity modes and transfers entanglement partially from detector C to the cavity modes, so that the entanglement between detectors degrades.
\begin{figure}
\begin{minipage}[t]{0.5\linewidth}
\centering
\includegraphics[width=3.0in,height=5.2cm]{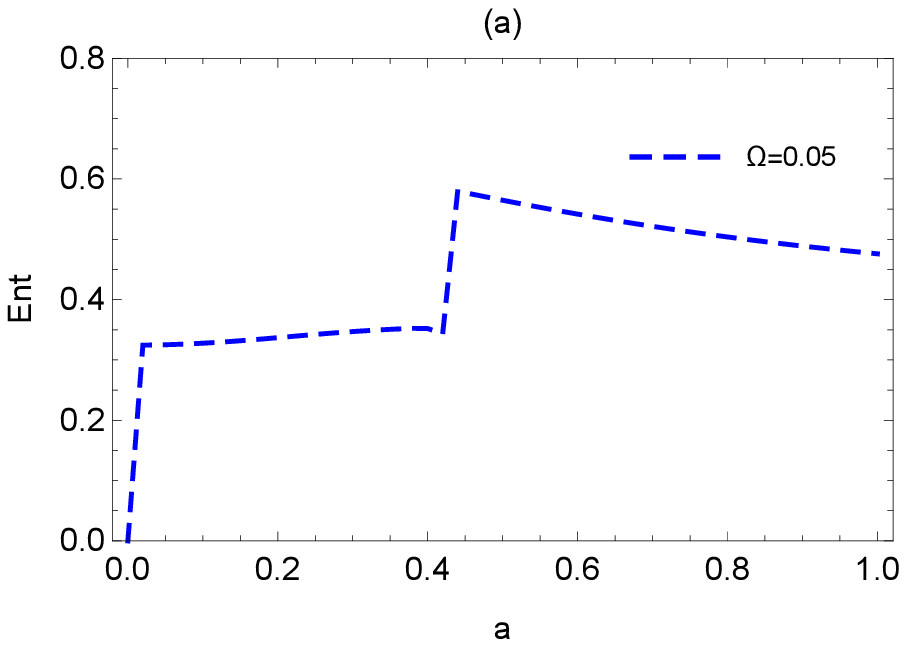}
\label{fig2a}
\end{minipage}%
\begin{minipage}[t]{0.5\linewidth}
\centering
\includegraphics[width=3.0in,height=5.2cm]{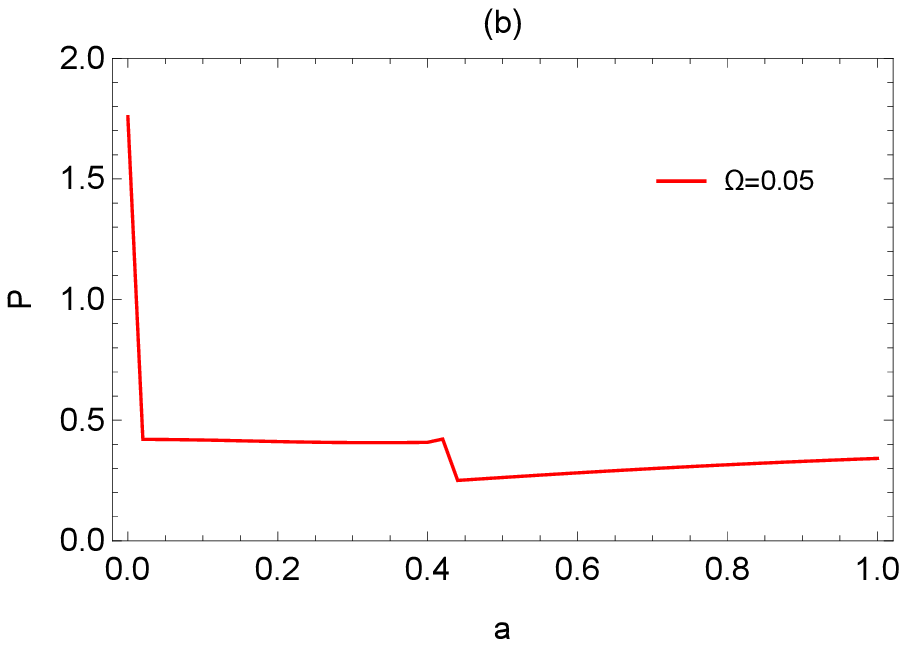}
\label{fig2b}
\end{minipage}%

\begin{minipage}[t]{0.5\linewidth}
\centering
\includegraphics[width=3.0in,height=5.2cm]{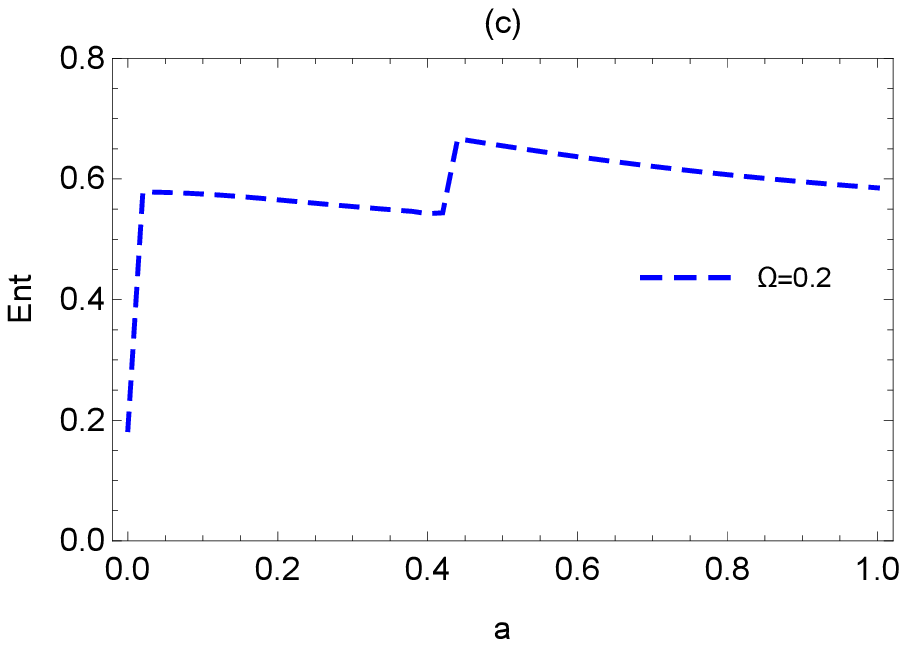}
\label{fig2c}
\end{minipage}%
\begin{minipage}[t]{0.5\linewidth}
\centering
\includegraphics[width=3.0in,height=5.2cm]{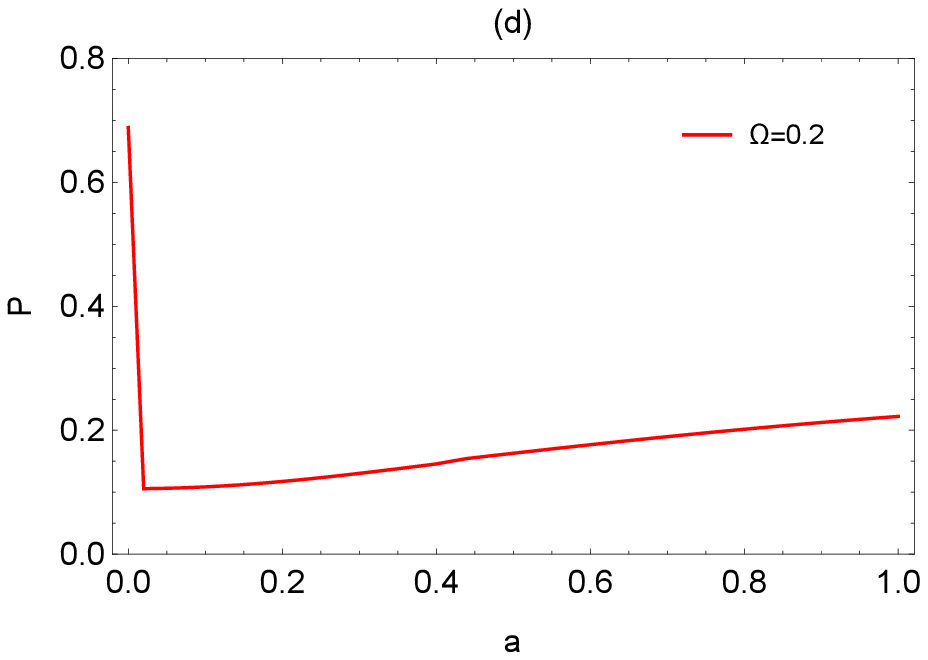}
\label{fig2d}
\end{minipage}%

\begin{minipage}[t]{0.5\linewidth}
\centering
\includegraphics[width=3.0in,height=5.2cm]{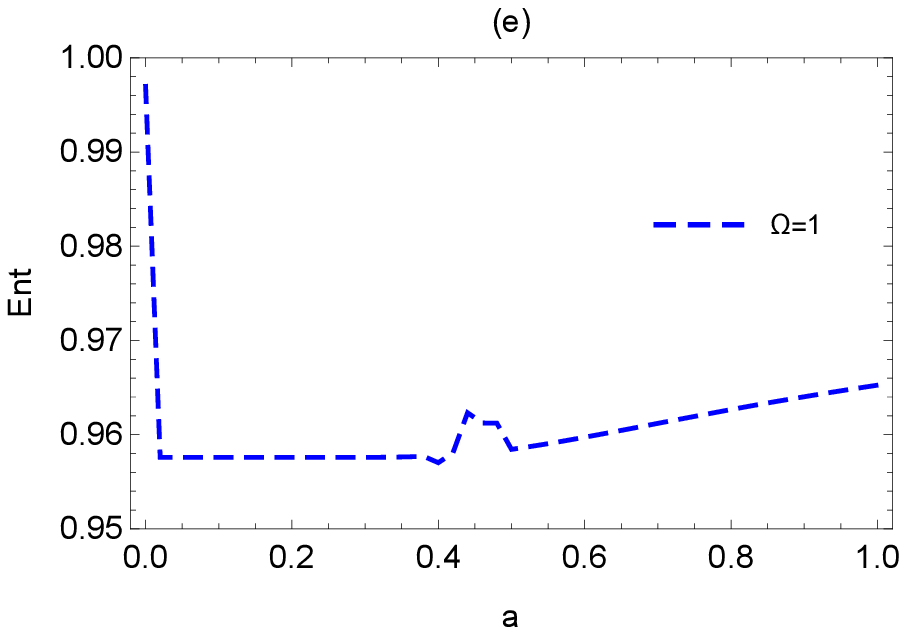}
\label{fig2e}
\end{minipage}%
\begin{minipage}[t]{0.5\linewidth}
\centering
\includegraphics[width=3.0in,height=5.2cm]{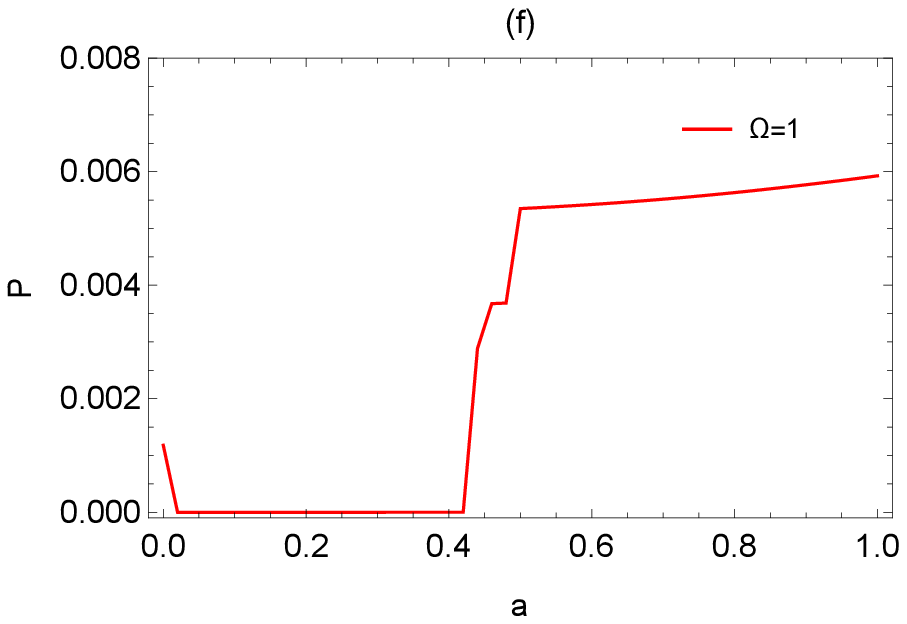}
\label{fig2f}
\end{minipage}%

\caption{Tripartite entanglement $E(\rho_{ABC_I})$ and the corresponding transition probability $P$ as functions of the acceleration $a$ for different the energy gap $\Omega$. The other parameters are fixed as $\sigma=5$, $L=200$ and $\alpha=\frac{1}{\sqrt{2}}$. }
\label{Fig2}
\end{figure}

In Fig.\ref{Fig2}, we plot the tripartite entanglement $E(\rho_{ABC_I})$ and the corresponding transition probability $P$, for the relatively larger interaction time ($\sigma=5$), as functions of the acceleration $a$ for different energy gap $\Omega$,
where other parameters are the same as in Fig.\ref{Fig1}. From the top row in Fig.\ref{Fig2}, we see that the anti-Unruh effect enhances entanglement (for about $a<0.45$) and Unruh effect decreases entanglement (for $a>0.45$). In the middle row in Fig.\ref{Fig2}, we see that for very small acceleration the anti-Unruh effect enhances entanglement; but later the Unruh effect both decreases and enhances entanglement. Finally, the last row in Fig.\ref{Fig2} shows that for very small acceleration the anti-Unruh decreases entanglement, and for about $a>0.42$ the Unruh effect can both enhance and decrease entanglement.

In addition, we point out that when the interaction timescale is far less than the timescale associated with the reciprocal of the detector's energy gap ($\sigma\ll 1/\Omega$), the anti-Unruh effect occurs for small acceleration (see Fig.\ref{Fig1}b and Fig.\ref{Fig2}b), which is consistent with the prediction of reference \cite{L20}. But in other cases (i.e., the condition $\sigma\ll 1/\Omega$ is not met or for larger acceleration), both Unruh and anti-Unruh can take place as seen in Fig.\ref{Fig1} and Fig.\ref{Fig2}.

From the analysis of Fig.\ref{Fig1} and Fig.\ref{Fig2} for one observer accelerating uniformly, we conclude that the increase and decrease in entanglement have no definite correspondence with the Unruh and anti-Unruh effects. Unruh effect can not only decrease but also enhance the tripartite entanglement between detectors; Also, anti-Unruh effect can not only enhance but also decrease the tripartite entanglement.

\begin{figure}
\begin{minipage}[t]{0.5\linewidth}
\centering
\includegraphics[width=3.0in,height=5.2cm]{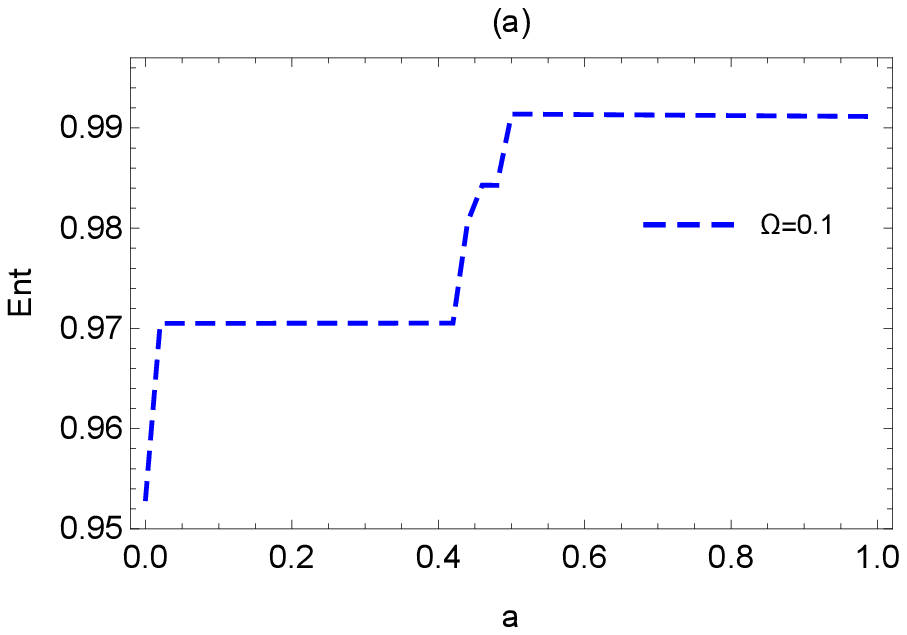}
\label{fig3a}
\end{minipage}%
\begin{minipage}[t]{0.5\linewidth}
\centering
\includegraphics[width=3.0in,height=5.2cm]{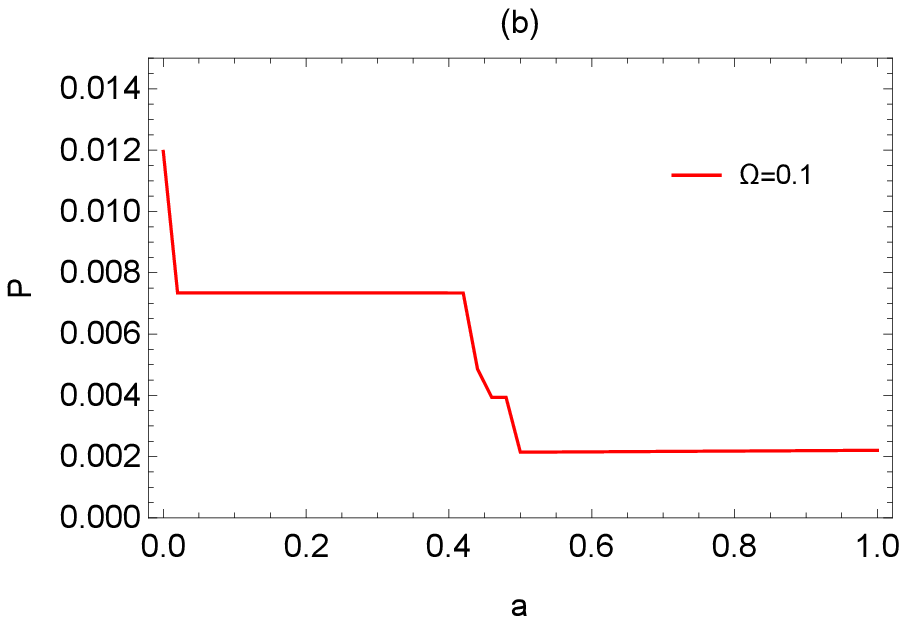}
\label{fig3b}
\end{minipage}%

\begin{minipage}[t]{0.5\linewidth}
\centering
\includegraphics[width=3.0in,height=5.2cm]{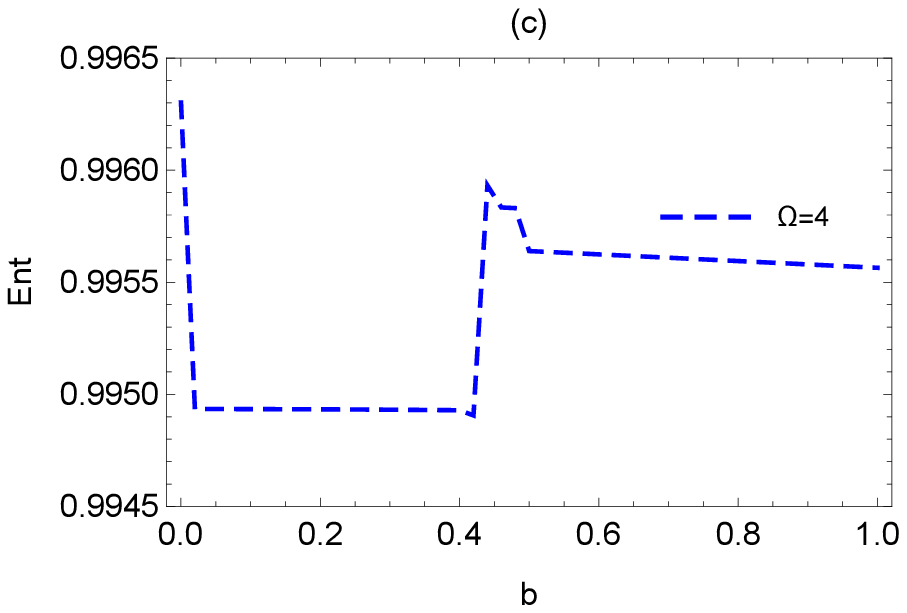}
\label{fig3c}
\end{minipage}%
\begin{minipage}[t]{0.5\linewidth}
\centering
\includegraphics[width=3.0in,height=5.2cm]{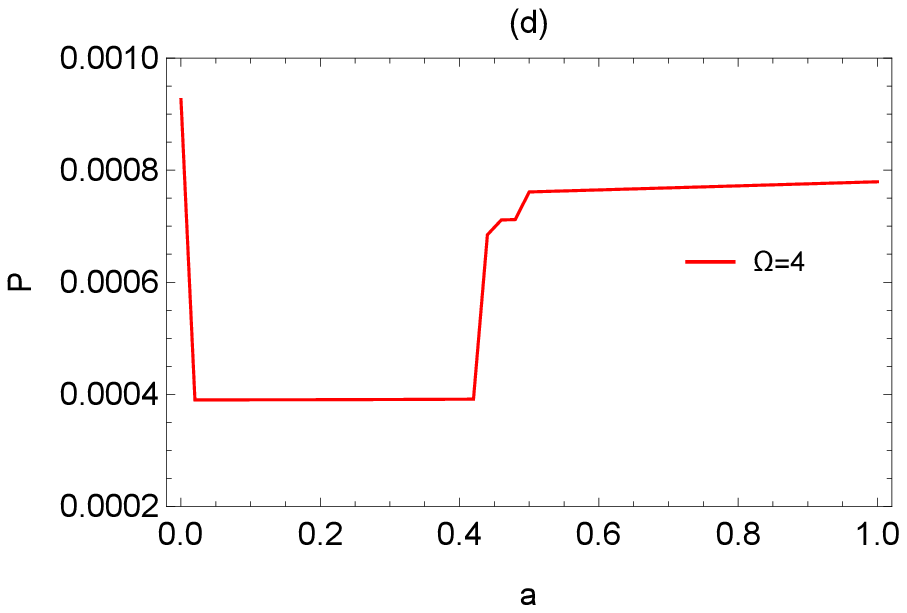}
\label{fig3d}
\end{minipage}%

\begin{minipage}[t]{0.5\linewidth}
\centering
\includegraphics[width=3.0in,height=5.2cm]{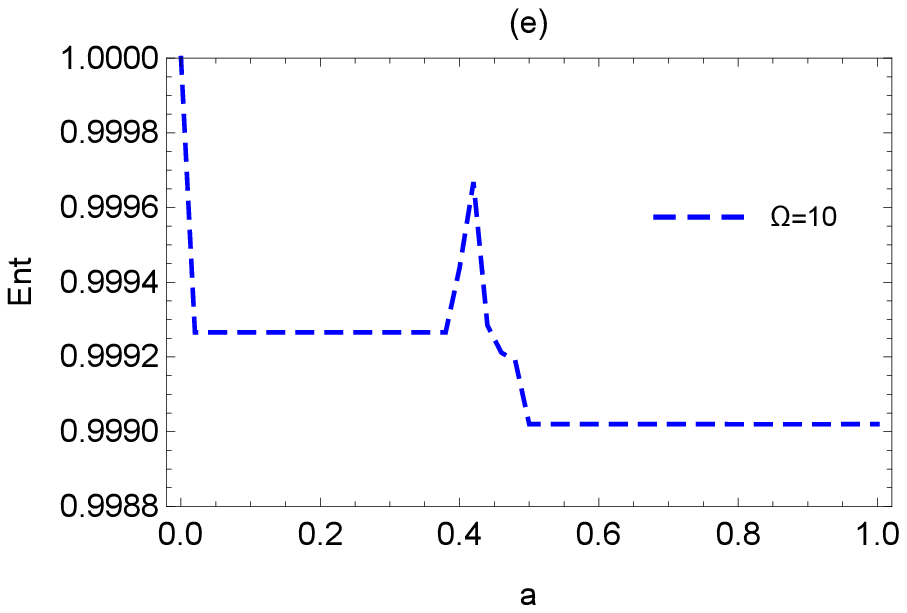}
\label{fig3e}
\end{minipage}%
\begin{minipage}[t]{0.5\linewidth}
\centering
\includegraphics[width=3.0in,height=5.2cm]{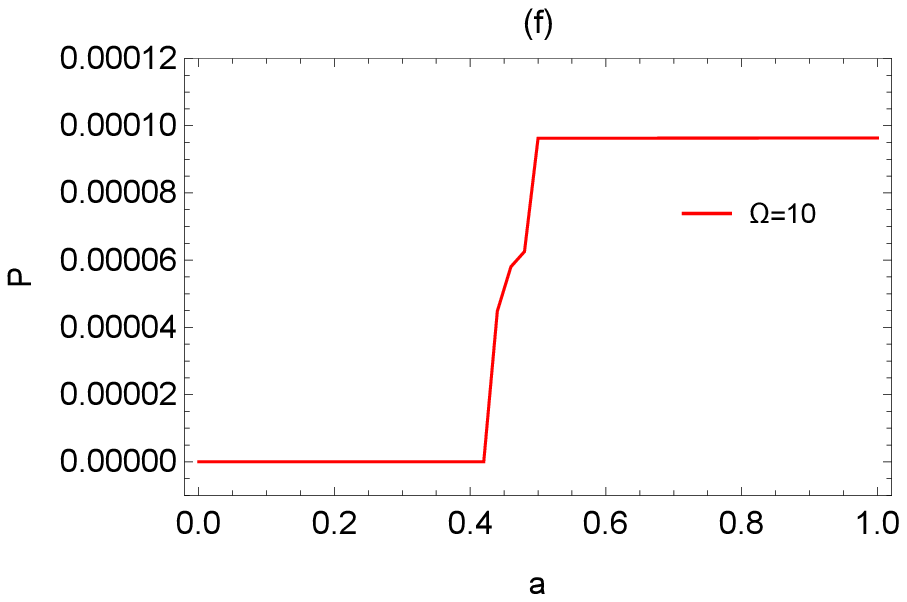}
\label{fig3f}
\end{minipage}%

\caption{Tripartite entanglement $E(\rho_{AB_IC_I})$  and the corresponding transition probability $P$ as functions of the acceleration $a$ for different the energy gap $\Omega$. The other parameters are fixed as $\sigma=0.4$, $L=200$ and $\alpha=\frac{1}{\sqrt{2}}$. }
\label{Fig3}
\end{figure}

\subsection{Bob and Charlie accelerating}
Now, we consider the case (ii), i.e., Alice remains stationary, while Bob and Charlie move with the same acceleration in cavities.
We want to know whether some new phenomena can be found in this case.

When both detectors (detectors B and C) move at a acceleration $a$, the initial state of Eq.(\ref{Q11}) becomes as
\begin{eqnarray}\label{Q16}
|\psi\rangle_{AB_IC_I}&=&\frac{\alpha}{1+|\eta_0|^2}|g_Ag_Bg_C0_A0_B0_C\rangle-
\frac{i\alpha\eta_0}{1+|\eta_0|^2}|g_Ae_Bg_C0_A1_B0_C\rangle \\ \nonumber
&-& \frac{i\alpha\eta_0}{1+|\eta_0|^2}|g_Ag_Be_C0_A0_B1_C\rangle-
\frac{\alpha\eta_0^2}{1+\eta_0^2}|g_Ae_Be_C0_A1_B1_C\rangle \\ \nonumber
&+&\frac{\beta}{1+|\eta_1|^2}|e_Ae_Be_C0_A0_B0_C\rangle+
\frac{i\beta\eta_1}{1+|\eta_1|^2}|e_Ag_Be_C0_A1_B0_C\rangle \\ \nonumber
&+&\frac{i\beta\eta_1}{1+|\eta_1|^2}|e_Ae_Bg_C0_A0_B1_C\rangle
-\frac{\beta\eta_1^2}{1+|\eta_1|^2}|e_Ag_Bg_C0_A1_B1_C\rangle \nonumber,
\end{eqnarray}
and the reduced density operator $\rho_{AB_IC_I}$ for the detectors reads
\begin{eqnarray}\label{Q17}
 \rho_{AB_IC_I}= \left(\!\!\begin{array}{cccccccc}
a_1 & 0 & 0 & 0 & 0 & 0 & 0 & z_1\\
 0 & a_2 & 0 & 0 & 0 & 0 & z_2& 0 \\
 0 & 0 & a_3 & 0 & 0 &  z_3 & 0 & 0 \\
0 & 0 & 0 & a_4 & z_4 & 0 & 0 & 0 \\
0 & 0 & 0 & z_4^* &  b_4 & 0 & 0 & 0\\
0 & 0 & z^*_3 & 0 & 0 & b_3 & 0 & 0\\
0 &z_2^* & 0 & 0 & 0 & 0 & b_2 & 0\\
z_1^* & 0 & 0 & 0 & 0 & 0 & 0 & b_1
\end{array}\!\!\right),
\end{eqnarray}
with the nonzero elements given by
\begin{eqnarray}\label{Q18}
&&a_1=\frac{\alpha^2}{(1+|\eta_0|^2)^2}, a_2=\frac{\alpha^2|\eta_0|^2}{(1+|\eta_0|^2)^2}, \\ \nonumber
&&a_3=\frac{\alpha^2|\eta_0|^2}{(1+|\eta_0|^2)^2}, a_4=\frac{\alpha^2|\eta_0|^4}{(1+|\eta_0|^2)^2}, \\ \nonumber
&&b_1=\frac{\beta^2}{(1+|\eta_1|^2)^2}, b_2=\frac{\beta^2|\eta_1|^2}{(1+|\eta_1|^2)^2}, \\ \nonumber
&&b_3=\frac{\beta^2|\eta_1|^2}{(1+|\eta_1|^2)^2}, b_4=\frac{\beta^2|\eta_1|^4}{(1+|\eta_1|^2)^2}, \\ \nonumber
&&z_1=\frac{\alpha\beta}{(1+|\eta_0|^2)(1+|\eta_1|^2)}, z_2=\frac{-\alpha\beta\eta_0\eta_1^*}{(1+|\eta_0|^2)(1+|\eta_1|^2)}, \\ \nonumber
&&z_3=\frac{-\alpha\beta\eta_0\eta_1^*}{(1+|\eta_0|^2)(1+|\eta_1|^2)}, z_4=\frac{\alpha\beta(\eta_0\eta_1^*)^2}{(1+|\eta_0|^2)(1+|\eta_1|^2)}. \\ \nonumber
\end{eqnarray}
Substituting these parameters into Eq.(\ref{Q4}), we obtain the tripartite entanglement of the detectors as
\begin{eqnarray}\label{Q19}
E(\rho_{AB_IC_I})=\max \bigg\{0, \frac{2\alpha\beta(1-2|\eta_0||\eta_1|-|\eta_0|^2|\eta_1|^2)}{(1+|\eta_0|^2)(1+|\eta_1|^2)},
\frac{2\alpha\beta(|\eta_0|^2|\eta_1|^2-2|\eta_0||\eta_1|-1)}{(1+|\eta_0|^2)(1+|\eta_1|^2)}\bigg\}.
\end{eqnarray}

\begin{figure}
\begin{minipage}[t]{0.5\linewidth}
\centering
\includegraphics[width=3.0in,height=5.2cm]{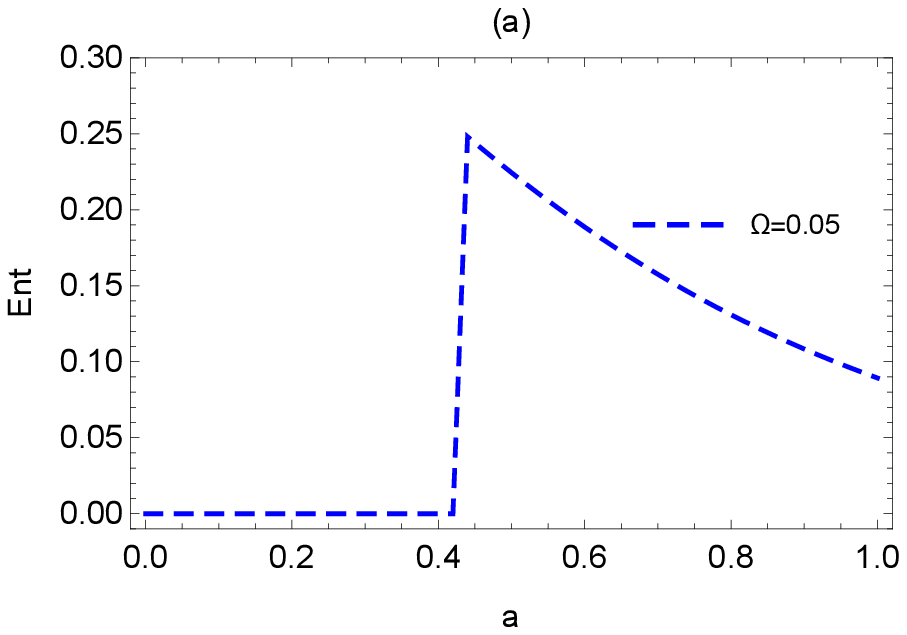}
\label{fig4a}
\end{minipage}%
\begin{minipage}[t]{0.5\linewidth}
\centering
\includegraphics[width=3.0in,height=5.2cm]{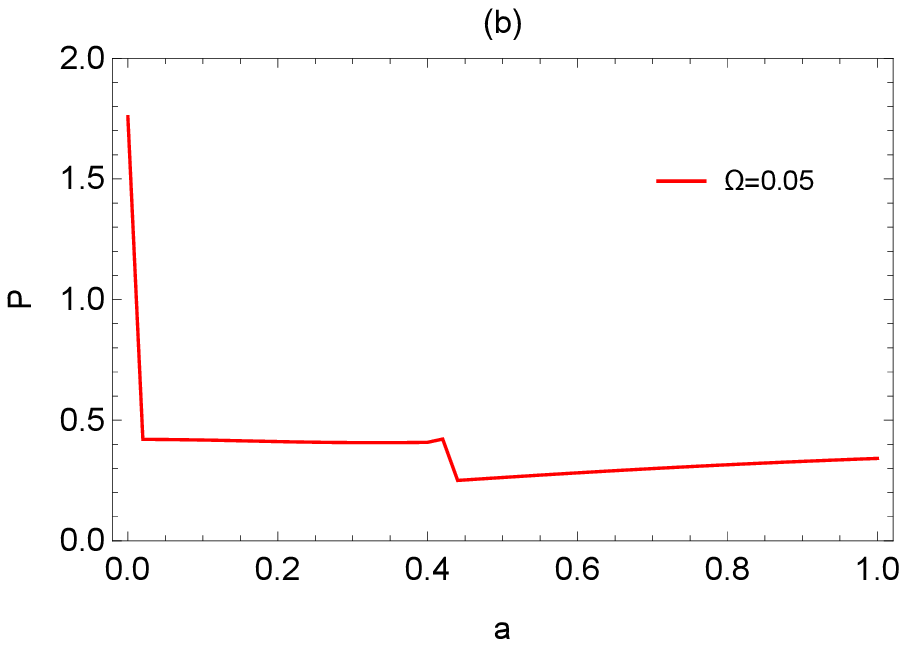}
\label{fig4b}
\end{minipage}%

\begin{minipage}[t]{0.5\linewidth}
\centering
\includegraphics[width=3.0in,height=5.2cm]{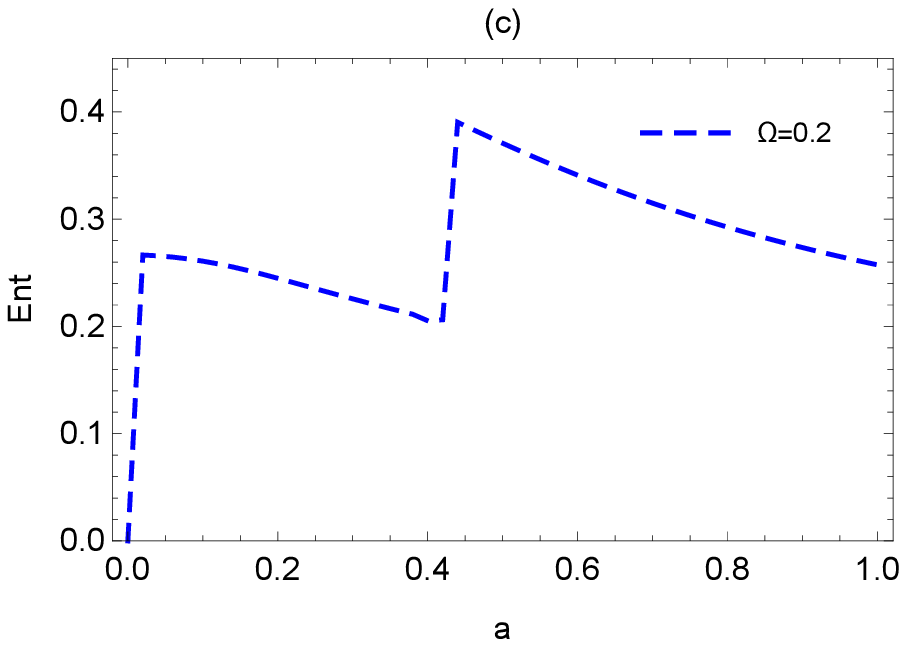}
\label{fig4c}
\end{minipage}%
\begin{minipage}[t]{0.5\linewidth}
\centering
\includegraphics[width=3.0in,height=5.2cm]{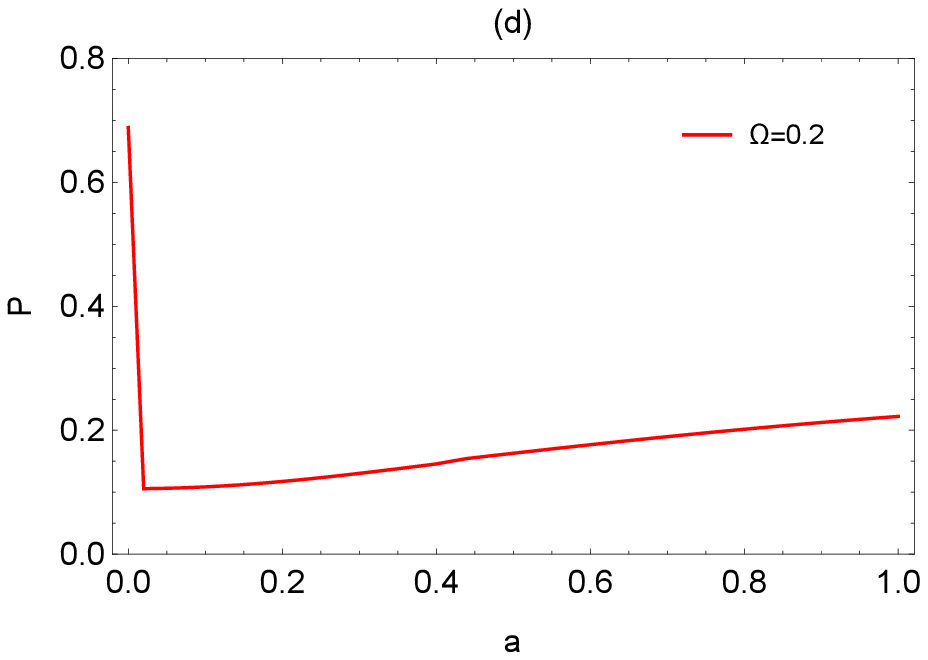}
\label{fig4d}
\end{minipage}%

\begin{minipage}[t]{0.5\linewidth}
\centering
\includegraphics[width=3.0in,height=5.2cm]{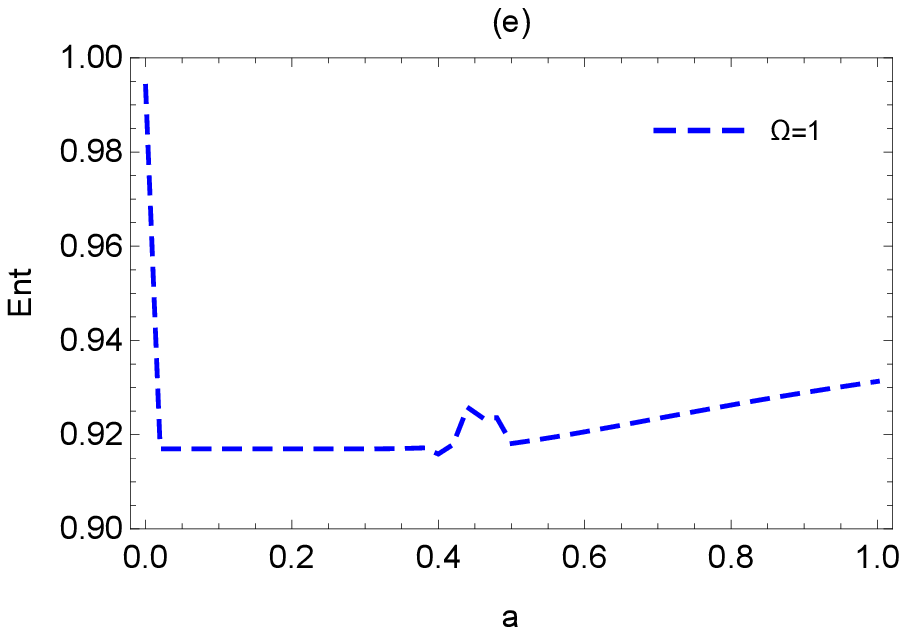}
\label{fig4e}
\end{minipage}%
\begin{minipage}[t]{0.5\linewidth}
\centering
\includegraphics[width=3.0in,height=5.2cm]{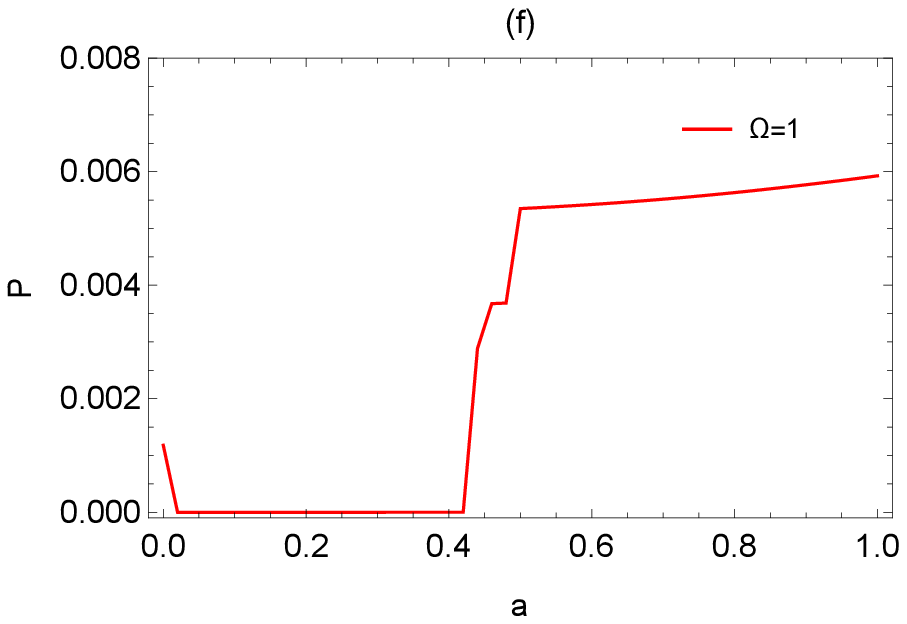}
\label{fig4f}
\end{minipage}%

\caption{Tripartite entanglement $E(\rho_{AB_IC_I})$  and the corresponding transition probability $P$ as functions of the acceleration $a$ for different the energy gap $\Omega$. The other parameters are fixed as $\sigma=5$, $L=200$ and $\alpha=\frac{1}{\sqrt{2}}$. }
\label{Fig4}
\end{figure}

In Fig.\ref{Fig3}, we plot the tripartite entanglement $E(\rho_{AB_IC_I})$ and the corresponding transition probability $P$ as functions of the acceleration $a$ for smaller interaction time $\sigma=0.4$. For the sake of comparison, the parameters are chosen the same as in Fig.\ref{Fig1}. As a whole, the evolution of entanglement versus acceleration is very similar to Fig.\ref{Fig1}. The remarkable difference may be that the corresponding values of $E(\rho_{AB_IC_I})$ in Fig.\ref{Fig3} is more less than $E(\rho_{ABC_I})$ in Fig.\ref{Fig1}. The reason for this may include two factors: One is due to the initial entanglement transfer between detector and vacuum cavity. When detector enters its vacuum cavity (even for zero acceleration), it couples with the vacuum cavity, leading to entanglement transfer from detector to cavity. Now there are two detectors coupling with their vacuum cavities, leading to more entanglement transfer from detectors to cavities. This can be seen by comparing Fig.\ref{Fig3}(a),(c) with Fig.\ref{Fig1}(a),(c). Another factor is due to the acceleration effect, now there are more observers (both Bob and Charlie) suffering from acceleration effect. This can be seen by comparing Fig.\ref{Fig3}(e) with Fig.\ref{Fig1}(e), where the initial entanglement for zero acceleration are both unity, but the entanglement for other accelerations fulfils $E(\rho_{AB_IC_I})<E(\rho_{ABC_I})$.

In Fig.\ref{Fig4}, we plot the tripartite entanglement $E(\rho_{AB_IC_I})$ and the corresponding transition probability $P$ as functions of the acceleration $a$ for relatively larger interaction time $\sigma=5$. For the sake of comparison, we take the same parameters as in Fig.\ref{Fig2}. Generally speaking, Fig.\ref{Fig4} is very similar to Fig.\ref{Fig2}, but has some differences. Firstly, the entanglement $E(\rho_{AB_IC_I})$ in Fig.\ref{Fig4} is more less than $E(\rho_{ABC_I})$ in Fig.\ref{Fig2}, because of the reason of more initial entanglement transfer and more number of detector's acceleration. Secondly, the entanglement in Fig.\ref{Fig4} (a) and (c) is zero for $a=0$, and then rises at some finite acceleration. We usually call the phenomenon that the entanglement increases from zero the ``entanglement sudden birth" (ESB). Physically, this ESB phenomenon originates from the combination of both the initial entanglement transfer and the entanglement recovery. When the detector enters into the vacuum cavity, it interacts with the vacuum cavity, and transfers entanglement completely from detector to cavity. The lost entanglement however can get back to the detector in the later acceleration. Therefore, ESB occurs. Compared with the case that one detector accelerates, ESB seems more easily to appear in the case that two detectors accelerate, as seen in Fig.\ref{Fig4} and Fig.\ref{Fig2}, because it is more easily to take place complete entanglement transfer when more detectors couple to vacuum cavities. Note that except for ESB, there is also a delay phenomenon in Fig.\ref{Fig4}(a), i.e., entanglement remains zero up to $a\simeq 0.42$.

\section{Extension to N-partite systems  \label{GSCDGE}}
Now, we extend the discussions from the tripartite systems to N-partite systems. Consider the N-partite GHZ-like state
\begin{eqnarray}\label{Q20}
|\psi\rangle_{i,\ldots,N}=\alpha|g\rangle^{\otimes N}+\beta|e\rangle^{\otimes N},
\end{eqnarray}
where the mode $i$ ($i = 1, 2, \ldots, N$) is observed by observer $O_i$.
Assume that $n $ ($n<N$) observers move at the same acceleration and the rest $N-n$ observers stay stationarily. Following the method for treating the tripartite systems, we obtain the expression of the N-partite entanglement
\begin{eqnarray}\label{Q21}
E(\rho_{N,n})&=&\max \bigg\{0, \frac{2\alpha\beta[1-\sum_{b=1}^n\mathbf{C}_{n}^b|\eta_0|^b|\eta_1|^b]}{(1+|\eta_0|^2)^{\frac{n}{2}}(1+|\eta_1|^2)^{\frac{n}{2}}}, \\ \nonumber
&&
\frac{2\alpha\beta[|\eta_0|^n|\eta_1|^n-1-\sum_{b=1}^{n-1}\mathbf{C}_n^b|\eta_0|^b|\eta_1|^b]}
{(1+|\eta_0|^2)^{\frac{n}{2}}(1+|\eta_1|^2)^{\frac{n}{2}}}\bigg\},
 \end{eqnarray}
with $\mathbf{C}_{n}^b=\frac{n!}{b!(n-b)!}$. Now besides the parameters involved in tripartite systems, $E(\rho_{N,n})$ also depends on the number $n$ of the accelerated observers. However, it is independent of the number of the total observers $N$.

\begin{figure}
\begin{minipage}[t]{0.5\linewidth}
\centering
\includegraphics[width=3.0in,height=5.2cm]{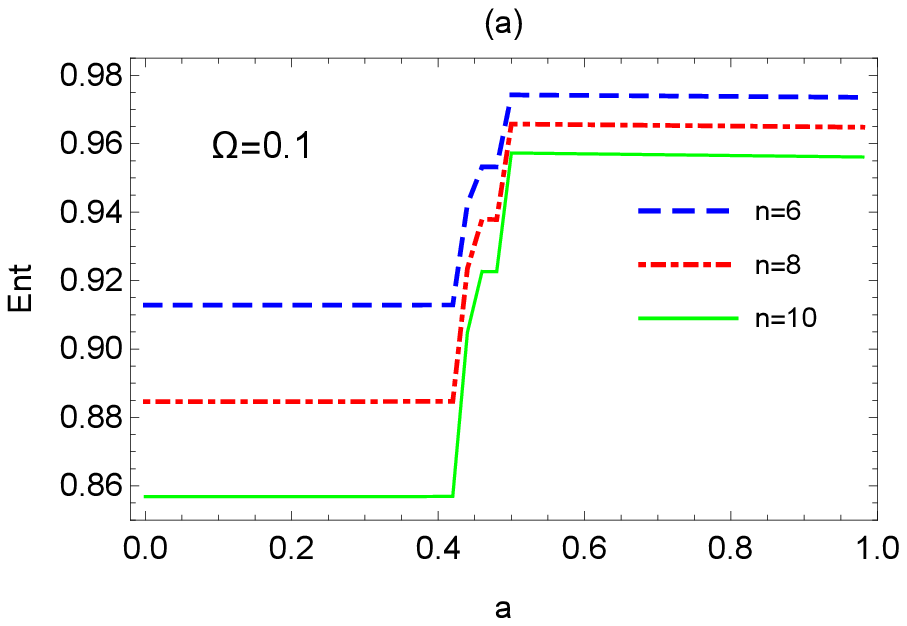}
\label{fig5a}
\end{minipage}%
\begin{minipage}[t]{0.5\linewidth}
\centering
\includegraphics[width=3.0in,height=5.2cm]{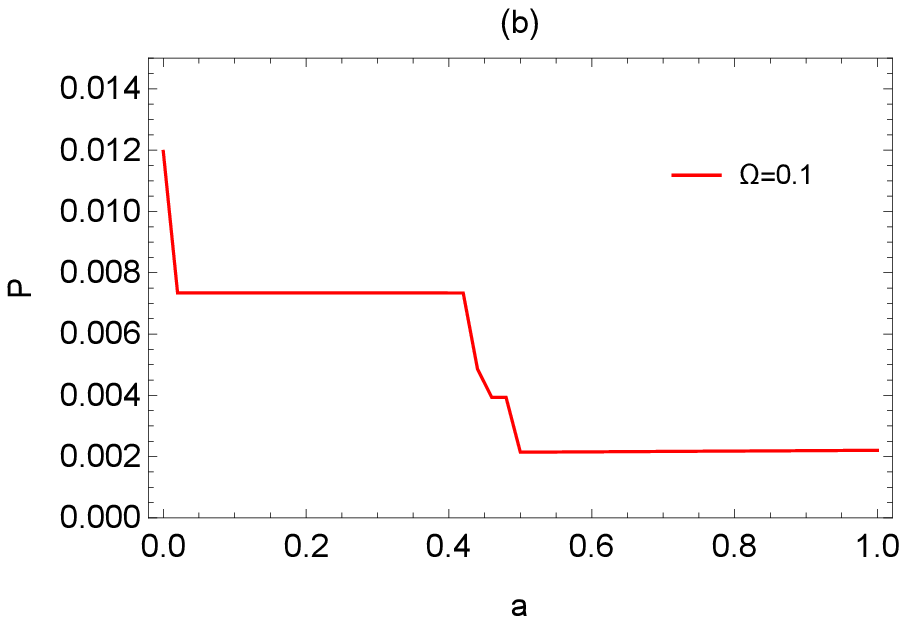}
\label{fig5b}
\end{minipage}%

\begin{minipage}[t]{0.5\linewidth}
\centering
\includegraphics[width=3.0in,height=5.2cm]{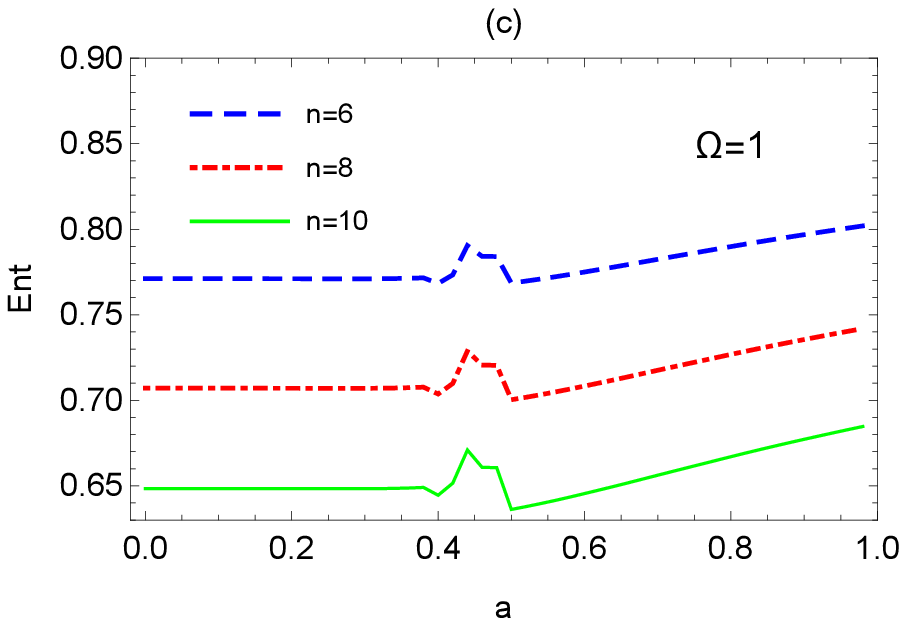}
\label{fig5a}
\end{minipage}%
\begin{minipage}[t]{0.5\linewidth}
\centering
\includegraphics[width=3.0in,height=5.2cm]{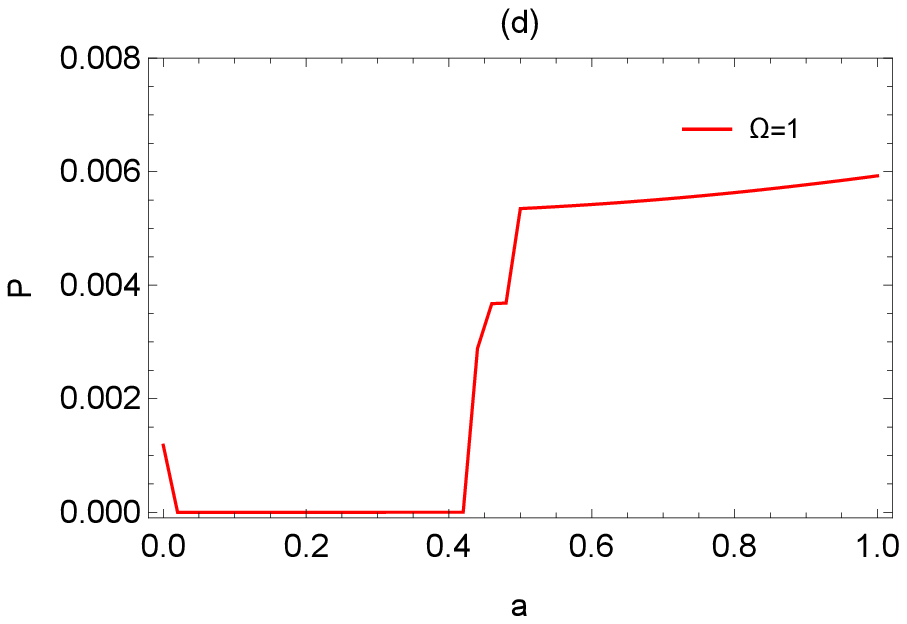}
\label{fig5b}
\end{minipage}%

\caption{N-partite entanglement $E(\rho_{N,n})$  and the corresponding transition probability $P$ as functions of the acceleration $a$. In (a)-(b)  $\sigma=0.4$ and in (c)-(d) $\sigma=5$.  The other parameters are fixed as $L=200$ and $\alpha=\frac{1}{\sqrt{2}}$. }
\label{Fig5}
\end{figure}

In Fig.\ref{Fig5}, we show the behavior of the N-partite entanglement and the corresponding transition probability $P$ as functions of acceleration $a$, for different $n$ and two kinds of interaction time $\sigma=0.4$ (top row) and $\sigma=5$ (bottom row).
It is shown that the N-partite entanglement $E(\rho_{N,n})$ decreases with the number $n$ of the accelerated observers. This is indeed consistent with the case of tripartite entanglement, where we saw that the entanglement for two-accelerated observers is less than the entanglement for one-accelerated observer. The reason for this also comes from two factors: the initial entanglement transfer between detectors and vacuum cavities, and the acceleration effect.

The top row ($\sigma=0.4$) in Fig.\ref{Fig5} shows that ant-Unruh effect enhances the N-partite entanglement, which is consistent with the previous result for bipartite entanglement \cite{L21}. But the bottom row ($\sigma=5$) shows different result. In the region about $a\in(0.4,0.5)$, the entanglement appears a peak but in which the corresponding transition probability increases monotonically, meaning that Unruh effect can both enhance and reduce N-partite entanglement.

\section{ Conclusion and discussion \label{GSCDGE}}
Based on the Unruh-DeWitt detector model, we have studied how acceleration affects tripartite entanglement between the detectors.
Two cases are considered: (i) One observer moves with a uniform acceleration, while the other two remain stationary; (ii) Two observers move at the same acceleration, while the other is stationary.
Different from the previous result that anti-Unruh effect enhances the entanglement between the detectors observed in bipartite systems \cite{L21}, we have found that the increase or decrease in entanglement has no definite correspondence with the Unruh and anti-Unruh effects. Both Unruh and anti-Unruh effects can enhance or reduce the tripartite entanglement between detectors. This is an interesting phenomenon, which is the main result for the Unruh and anti-Unruh effects in this paper. In addition, we have seen the acceleration can either enhance or degrade harvested entanglement, although this effect of acceleration has been shown in two detector systems previously, the possible relevance of the anti-Unruh effect was not considered \cite{LAZ22,LAZ23,LAZ24}.

Besides the main result, we have also found some by-products. Firstly, given the same parameters, the tripartite entanglement decreases with the number $n$ of the accelerated detectors (This result is also valid for the N-partite entanglement!). The reason for this comes from two factors. One is due to the initial entanglement transfer between detector and vacuum cavity. When a detector enters its vacuum cavity, it couples with the vacuum cavity, leading to entanglement transfer from detector to cavity. Two-detector coupling can transfer more entanglement than one-detector coupling. Another factor is due to the acceleration effect. Two-detector accelerating suffers from more Fulling-Unruh radiation than one-detector accelerating. Secondly, we have observed the phenomena of ESB and delay. It comes from the combination of the initial entanglement transfer and the later entanglement recovery. During the coupling of the detector with its vacuum cavity, entanglement is transferred completely to the vacuum cavity. Afterwards, when the detector accelerates uniformly, the entanglement transfers back from the cavity to the detector. In this way, ESB takes place. Some times, the recovery of entanglement needs a finite acceleration, leading to the delay of ESB. Obviously, the case that two detectors accelerate is more easily to appear ESB than the case that one detector accelerates, as we have observed in the text.
From the relativistic regime, this ESB behavior denotes the production of entanglement via acceleration effect.

We have also extended the discussion from the tripartite to N-partite systems. Besides the similar results as in tripartite systems, some new results have been found.
The N-partite entanglement between detectors depends only on the number $n$ of the accelerated detectors, not on the total number $N$ of the detectors. When other parameters are fixed, the N-partite entanglement between detectors reduces with $n$. The reason also comes from the entanglement transfer and the acceleration effect, as in the tripartite entanglement.

In nature, the Unruh effect is established for quantum fields. The action that the Fulling-Unruh radiation superimposes on the quantum fields behaves like a noise, which leads to the irreversibly loss of quantum information encoding in the quantum fields \cite{L7,L8,L9}.
However, the problem we here consider is the entanglement between detectors (two-level atoms) moving weakly (i.e. perturbatively) toward equilibrium with the fields rather than between quantum fields themselves. Now the Unruh and anti-Unruh effects are defined through the detector's excitation ($|g\rangle\rightarrow|e\rangle$) and de-excitation ($|e\rangle\rightarrow|g\rangle$). The measure of entanglement is independent of the nature of
$|g\rangle$ and $|e\rangle$. For any two-qubit entangled state, after the exchange of ground state $|g\rangle$ and excited state $|e\rangle$, the entanglement remains unchanged. Consider a two-detector system, in which one detector is stationary and the other accelerates uniformly in a vacuum cavity. Let $|\Psi(a_{1})\rangle$ and $|\Psi(a_{2})\rangle$ denote the two entangled states of the two-detector system with accelerations $a_{1}$ and $a_{2}$ (assume $a_{2}>a_{1}$), and assume that the entanglement of $|\Psi(a_{2})\rangle$ is less than the entanglement of $|\Psi(a_{1})\rangle$. If the detector is excited more stronger in acceleration $a_{2}$ than in acceleration $a_{1}$, then we say that Unruh effect reduces entanglement between the detectors, and the process is described as $|\Psi(a_{1})\rangle\stackrel{U}{\longrightarrow}|\Psi(a_{2})\rangle$. Now by exchanging basis $|g\rangle\leftrightarrow|e\rangle$, we obtain states $|\widetilde{\Psi}(a_{1})\rangle$ and $|\widetilde{\Psi}(a_{2})\rangle$. As basis exchange does not alter entanglement of quantum states, thus the process from $|\widetilde{\Psi}(a_{1})\rangle$ to $|\widetilde{\Psi}(a_{2})\rangle$ also leads to reduction of entanglement. But in this time, the detector is de-excited, i.e., it is an anti-Unruh process. Thus we get $|\widetilde{\Psi}(a_{1})\rangle\stackrel{AU}{\longrightarrow}|\widetilde{\Psi}(a_{2})\rangle$, where the entanglement reduces. In this way, from the hypothesis that Unruh effect reduces entanglement, we deduce the result that anti-Unruh effect also reduces entanglement. Similarly, we can also from the hypothesis that anti-Unruh effect enhances entanglement deduce the result that Unruh effect enhances entanglement. This demonstrates our result that the increase and decrease in entanglement have no definite correspondence with the Unruh and anti-Unruh effects. This explanation though is very simple, but has not been obvious to previous authors on this topic.
We believe our result is a valuable correction to the impression given in some earlier work in this field which suggested otherwise. Of course, further research is required to demonstrate the validity of this inference.

\begin{acknowledgments}
This work is supported by the National Natural
Science Foundation of China (Grant Nos.1217050862, 11275064), and 2021BSL013.

\end{acknowledgments}

\textbf{Conflict of Interest}

The authors declare no conflict of interest.

\textbf{Data availability statement}

All data that support the findings of this study are included within the article.


\end{document}